\documentclass[conference,a4paper,10pt]{worldcomp}

\usepackage[hmargin=.75in,vmargin=1in]{geometry}
\usepackage[american]{babel}
\usepackage[T1]{fontenc}
\usepackage{times}
\usepackage{caption}

\newtheorem{rem}{Remarks}[section]
\newtheorem{theorem}{Theorem}

\usepackage{textcomp}
\usepackage{epsfig,graphicx}
\usepackage{xcolor}
\usepackage{amsfonts,amsmath,amssymb}
\usepackage{fixltx2e} 
\usepackage{booktabs}

\usepackage[dvipdfm,bookmarks=true,bookmarksnumbered=true,%
bookmarkstype=toc=true,colorlinks=true,urlcolor=black,%
linkcolor=black,citecolor=black,linktocpage=true,%
pdftitle={Bayesian Forecasting of WWW Traffic on the Time Varying Poisson Model},%
pdfsubject={The 2009 International Conference on Parallel and Distributed Processing Techniques and Applications (PDPTA'09), Las Vegas, Nevada, USA in July, 2009.},%
pdfauthor={Daiki Koizumi (dkoizumi@m.ieice.org)},%
pdfkeywords={World Wide Web (WWW) traffic, traffic engineering, statistical decision theory, time varying Poisson distribution, long-range dependence (LRD)}]{hyperref}

\title{\bf Bayesian Forecasting of WWW Traffic \\on the Time Varying Poisson Model}           

\author{
{\bfseries Daiki Koizumi$^1$, Toshiyasu Matsushima$^2$, and Shigeichi Hirasawa$^1$}\\
$^1$Research Institute for Science and Engineering, Waseda University, Tokyo, Japan\\
$^2$Department of Applied Mathematics, Waseda University, Tokyo, Japan\\
}

\begin{document}

\maketitle                        

\begin{abstract}

Traffic forecasting from past observed traffic data with small calculation complexity is one of important problems for planning of servers and networks. Focusing on World Wide Web (WWW) traffic as fundamental investigation, this paper would deal with Bayesian forecasting of network traffic on the time varying Poisson model from a viewpoint from statistical decision theory. Under this model, we would show that the estimated forecasting value is obtained by simple arithmetic calculation and expresses real WWW traffic well from both theoretical and empirical points of view.

\end{abstract}

\vspace{1em}


\section{Introduction}
\label{intro}

Under network environment such as Internet, planning of servers and networks is one of important problems for stable operation. It is often typical situation that administrators analyze logs on their servers and networks. They may frequently look into result of log analysis software where these tools usually have some functions to periodically summarize logs. For example, webalizer \cite{Barrett} and analog \cite{Turner} etc.\, have been widely used among World Wide Web (WWW) server administrators or users for long years. These tools usually summarize the logs by counting hourly, daily, and monthly numbers of hits, files, and pages etc. Administrators would often make their operation plans with combination of their experience and intuition from these logs. In this case, traffic forecasting rule is not clearly formulated and those summarized logs remain in the field of \textit{descriptive statistics} from the statistical point of view.

On the other hand, researchers in the field of traffic engineering have been suggesting a lot of analysis models. Probabilistic approach is one of viewpoints in this field. It is wide-spread fact that the stationary Poisson distribution is not always suitable for Internet traffic because of its nature of non-stationality \cite{Paxon} \cite{Karagiannis} and long-range dependence (LRD) \cite{Leland} \cite{Paxon} etc. Therefore desirable conditions of good traffic models are to have structures to express such nature at least. Furthermore, another requirement of models is to have a structure of traffic forecasting. For this point, parameter estimation is often performed at first under assumption of the stationarity \cite{Scherrer}, then the estimated parameter is substituted for the parameter of model. This approach has been wide-spread in the field of \textit{inferential statistics} from the statistical point of view.

However, substituting the estimated parameter as a constant for the model's parameter is not always suitable especially on forecasting problems. This is because there is often no guarantee that the assumptions under the parameter estimation of the model always hold for future unknown data set. Bayesian approach \cite{Berger}\cite{Bernardo} is one of alternatives for this point. In Bayesian approach, a probability distribution of parameter is assumed as the prior distribution. If new data is observed, then the Bayes theorem updates the prior distribution of parameter to the posterior distribution and then forecasts the posterior distribution of data. Recently, this approach has been widely applied to many forecasting problems especially in the field of information technologies and bioinformatics etc. In order to take Bayesian approach, \textit{statistical decision theory} is an important theoretical framework from the statistical point of view.

Taking the above factors into account, this paper would deal with Bayesian forecasting of WWW traffic on the non-stationary i.e. time varying Poisson model. Bayesian forecasting on time varying parameter model has been proposed in \cite{Smith} by defining certain class of parameter transformation function. However, it has not yet been discussed about any predictive estimator nor definite transformation function of parameter \cite{Smith}. This paper would clearly define a random-walking type of transformation function of parameter to obtain the Bayes optimal prediction for WWW traffic. Then its effectiveness would be evaluated with real WWW traffic data. In this model, time varying degree is caught by a real valued constant $k \,(0 < k \leq 1)$ and this constant would play an important role throughout this paper. Another feature is that the traffic forecasting value is obtained by simple arithmetic calculations under known $k$. In general, the Bayes theorem often results in large calculation costs. However, certain combination of parameter distribution and its transformation function solves this problem. We believe that this point can be helpful not only for theoretical calculation cost but also for real implementation on WWW log analysis tools \cite{Barrett} \cite{Turner}.

The rest of this paper is organized as the followings. Section \ref{Model Def} gives some definitions and explanations of the forecasting model with time varying Poisson distribution.
Section \ref{Simulation} shows some analysis examples of real WWW traffic data  to validate this paper's approach and Section \ref{Discussion} gives{\large } their discussions. Finally, Section \ref{Conclusion} concludes this paper.


\section{The Time Varying Poisson Model}
\label{Model Def}

\subsection{Definitions}
\label{Definiton of the Proposed Model}

Suppose $X$ is a discrete random variable and $p(x)=Pr \{ X=x \}$ is the probability distribution where real number $x$ is each element of space of $X$. Throughout this paper, the probability distribution of $X$ is assumed to depend on real valued parameter $\theta \in \Theta$. The true parameter $\theta^{*} \in \Theta$ is unknown, however, the probability distribution of parameter $p(\theta)$ is assumed to be known. Hereafter the probability distribution of $X$ under parameter $\theta$ is simply denoted as $p(x \bigm| \theta)$.
\begin{figure}[h]
\begin{center}
\includegraphics[width=\columnwidth]{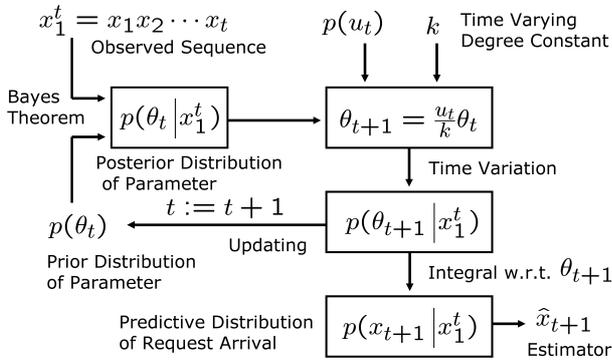}
\caption{Overview of the inferential process.}
\label{Model}
\end{center}
\end{figure}

Let $t=1,2,\cdots$ be discrete time and $x_{t}=0,1,\cdots$ be number of WWW request arrivals at time $t$, respectively. This paper focuses on $x_{t}$ for traffic analysis by assuming probability distribution $p(x_{t} \bigm | \theta_{t})$ where $\theta_{t}>0$ is a time varying density parameter at time $t$. The time varying Poisson model takes sequence of $x_{1}^{t}=x_{1}x_{2}\cdots x_{t}$ as input and calculates $\hat{x}_{t+1}$ as an output estimator where the prior distribution of parameter $p(\theta_{t})$ and time variation rule of $\theta_{t}$ are known. The overview of the inferential process is depicted in Figure \ref{Model}. 

In Figure \ref{Model}, $x_{t}$ is assumed to be the Poisson distribution with a time varying density parameter $\theta_{t}$ as follows:

For $x_{t}=0,1,2,\cdots$,
\begin{IEEEeqnarray}{Cc}
\label{poisson}
p(x_{t}\bigm|\theta_{t}) &= \frac{\exp\left(-\theta_{t}\right)}{x_{t}!}\left(\theta_{t}\right)^{x_{t}} ,
\end{IEEEeqnarray}
where $\theta_{t}>0$ is a time varying density parameter.

For parameter $\theta_{t}$, the following time varying model is assumed:

For $\theta_{t+1}, \theta_{t}>0$,\
\begin{IEEEeqnarray}{Cc}
\label{theta_update}
\theta_{t+1}&=\frac{u_{t}}{k}\theta_{t} ,
\end{IEEEeqnarray}
where $k$ is a constant such that $0<k\leq1$, and $0< u_{t} <1$ is a continuous random variable which is conditionally independent from $\theta_{t}$. (\ref{theta_update}) represents a transformation of $\theta_{t+1}$ from both $\theta_{t}$ and $u_{t}$ under a known constant $k$. This transformation is regarded as a kind of random-walk.

Furthermore, the initial random variables of $\theta_{1}$ and $u_{1}$ are from the Gamma and Beta distributions, respectively. The followings give their definitions:

For $\theta_{1}>0$,
\begin{IEEEeqnarray}{lL}
\label{theta}
p(\theta_{1}\bigm|\alpha_{1},\beta_{1}) &= \frac{\left(\beta_{1}\right)^{\alpha_{1}}}{\Gamma\left(\alpha_{1}\right)}\exp \left(-\theta_{1}\beta_{1}\right)\left(\theta_{1}\right)^{\left(\alpha_{1}\right)-1} ,
\end{IEEEeqnarray}
where $\alpha_{1}>0,\ \beta_{1}>0$ are parameters of the Gamma distribution.

In (\ref{theta}), $\Gamma\left(x\right)$ is the Gamma function defined below:
\begin{IEEEeqnarray}{Cc}
\Gamma\left(x\right) &= \int^{\infty}_{0} y^{x-1}\exp\left(-y\right)dy ,
\end{IEEEeqnarray}

where $x>0$.

For $0< u_{1} <1$,
\begin{IEEEeqnarray}{Ll}
\lefteqn{p\left(u_{1}|k\alpha_{1},\left(1-k\right)\alpha_{1}\right)} \IEEEnonumber \\
\label{u}
&= \frac{\Gamma \left( \alpha_{1} \right)}{\Gamma \! \left( k\alpha_{1}\right) \! \Gamma [\left(1 \! - \! k \right)\alpha_{1}]} \! \left(u_{1}\right)^{\left(k\alpha_{1}\right)-1}\! \left( 1 \! - \! u_{1}\right)^{[\left(1-k\right)\alpha_{1}]-1} , \IEEEnonumber \\
&
\end{IEEEeqnarray}
where $k\alpha_{1}>0,\ \left(1-k \right)\alpha_{1}>0$ are parameters of the Beta distribution.

In (\ref{theta}) and (\ref{u}), two random variables $\theta_{1}$ and $u_{1}$ have a real value $\alpha_{1}>0$ as their parameters in common. This means that $\theta_{1}$ and $u_{1}$ are conditionally independent. 

\begin{rem}
\label{remark1 of model}
In (\ref{theta_update}), a constant $0 < k \leq 1$ expresses time varying degree of $\theta_{t}$. If $k=1$, (\ref{theta_update}) simply becomes $\theta_{t+1}=u_{t}\theta_{t}$. In this case, $\theta_{t+1}$ does not vary since the variance of $u_{t}$ , which equals to $k(1-k)/(\alpha_{t}+1)$ according to the nature of Beta distribution, becomes zero. This means that the Poisson distribution of $x_{t}$ in (\ref{poisson}) is stationary.

If $k < 1$, on the other hand, $\theta_{t+1}$ varies depending on the previous $\theta_{t}$ which expresses the time varying Poisson model. If $k=0.5$, the time varying degree of $\theta_{t}$ becomes maximum since the variance of $u_{t}$ takes the maximum value. Thus the proposed model defined in (\ref{poisson})--(\ref{u}) includes a classical stationary Poisson distribution as a special case if $k=1$. Moreover, $k$ plays another important role which would be explained in \textit{Remarks \ref{remark2 of model}}.
\end{rem}

\subsection{Updating Density Parameter $\theta_{t}$}
\label{updating parameter}

As shown in Figure \ref{Model}, the parameter updating rule from $\theta_{t}$ to $\theta_{t+1}$ consists of Bayes theorem and time variation.
In the followings, this subsection \ref{updating parameter} is divided into three parts. The first \ref{posterior prob. of theta} describes the updating rule by Bayes theorem for $t\geq2$, the second \ref{initial condition of theta} describes the initial condition of the prior distribution $p\left(\theta_{1}\right)$ for $t=1$, and the last \ref{time variation of theta} describes general time variation rule which is mainly discussed in \cite{Smith}.

\subsubsection{The Posterior Distribution of Density Parameter}
\label{posterior prob. of theta}


Suppose that sequence $x_{1}^{t-1}$ is already observed where $t\geq2$ and the prior distribution of parameter $p\left(\theta_{t}\bigm|\alpha_{t},\beta_{t},x_{1}^{t-1}\right)$ is defined as the Gamma distribution in (\ref{theta}) for $t \geq 2$. If new $x_{t}$ is observed, the posterior distribution of $p\left(\theta_{t}\bigm|x_{1}^{t}\right)$ by the Bayes theorem is obtained as the following: 
\begin{IEEEeqnarray}{Ll}
\lefteqn{p\left(\theta_{t}\bigm|x_{1}^{t}\right)} \IEEEnonumber \\
 &= \frac{p(x_{t}\bigm|\theta_{t})p(\theta_{t}\bigm|\alpha_{t},\beta_{t},x_{1}^{t-1})}{\int_{0}^{\infty}p(x_{t}\bigm|\theta_{t})p(\theta_{t}\bigm|\alpha_{t},\beta_{t},x_{1}^{t-1})d\theta_{t}} \\
\label{before2}
 &= \frac{\left(\beta_{t} \! + \! 1 \right)^{\alpha_{t}+x_{t}}}{\Gamma\left(\alpha_{t} \! + \! x_{t}\right)} \! \exp[-\theta_{t}(\beta_{t} \! + \! 1)] \! \left(\theta_{t}\right)^{\left(\alpha_{t}+x_{t}\right)-1}\,,
\end{IEEEeqnarray}
where $t \geq 2$.

(\ref{before2}) means that the posterior distribution $p\left( \theta_{t} \bigm| x_{1}^{t} \right)$ is also Gamma with parameter $\left( \alpha_{t}+x_{t} \right)$ and $\left( \beta_{t}+1 \right )$ . 

\subsubsection{Initial Condition of Prior Distribution of Density Parameter}
\label{initial condition of theta}

For the initial prior distribution of $p\left(\theta_{1}\right)$, this paper assumes no anomalies for WWW traffic. In Bayesian context, {\it non-informative prior} \cite{Berger}\cite{Bernardo} can be considered as the following:
\begin{IEEEeqnarray}{Cc}
\label{initial prior}
p\left(\theta_{1}\right) & \propto \frac{1}{\theta_{1}}\,.
\end{IEEEeqnarray}

The above distribution can be formulated by the Gamma distribution with parameters $\alpha_{1}>0, \beta_{1}=1$ in (\ref{theta}). Therefore, the general posterior updating form in (\ref{before2}) can also be applied for $t=1$. Thus (\ref{before2}) holds for any $t \geq 1$. This means that the posterior distribution $p \left(\theta_{t} \bigm| x_{1}^{t} \right)$ can be simply calculated for any $t \geq 1$ by considering two parameters $\alpha_{t}, \beta_{t}$ on the Gamma distribution if the initial prior distribution of (\ref{initial prior}) is assumed. This is the nature of {\it conjugate prior} \cite{Berger} \cite{Bernardo} between the Poisson and Gamma distributions. This nature contributes drastic reduction of calculation complexity under large $t$.


\subsubsection{Time Variation of Density Parameter}
\label{time variation of theta}

To obtain $p\left(\theta_{t+1}\bigm|x_{1}^{t}\right)$, a time variation of density parameter defined in (\ref{theta_update}) is used. This is actually a transformation of random variables among $u_{t}, \theta_{t}$, and constant $k$. Even if a transformation is newly defined after Bayes theorem, the distribution family of density parameter remains same as that of the conjugate prior distribution under certain class of transformations. Such class has been discussed in \cite{Smith} as the following.

\begin{theorem}[Simple Power Steady Model \cite{Smith}]
Suppose a parameter distribution $p(\theta_{t})$ is in the linear expanding family. Let $T$ be a transformation function of parameter $\theta_{t}$. If $T$ satisfies the following condition, then the parameter distribution remains same family of distribution not depending on $t$ and the forecasting model is called \textit{Simple Power Steady Model (S.P.S.M.)} with respect to $T$:
\begin{equation}
\label{smith}
T(x)= \Psi x^{k}, \,\, \Psi>0,\, 0<k<1\,.
\end{equation}
\end{theorem}

If $T$ satisfies (\ref{smith}) and is one-to-one mapping, the model can also be S.P.S.M.\cite{Smith}. However, it has not yet been discussed about any definite transformation function which is required to obtain the forecasting estimator. In fact, the time varying parameter model defined in (\ref{theta}) is one-to-one mapping since Jacobian of (\ref{theta}) is easily proved to be non-zero. Thus the model in this paper is actually included in S.P.S.M. The forecasting estimator would be derived in the next subsection \ref{output}.

Under S.P.S.M. in this paper, the transformed distribution of $\theta_{t+1}$ now becomes:
\begin{IEEEeqnarray}{Ll}
\lefteqn{p \left(\theta_{t+1} \Bigm|x_{1}^{t} \right)} \IEEEnonumber \\
\label{after1}
& = p \left(\frac{u_{t}}{k}\theta_{t} \Bigm| x_{1}^{t} \right) \\
\label{after2}
& = \frac{\left[k \left( \beta_{t} + 1 \right ) \right]^{k \left( \alpha_{t}+x_{t}\right)}}{\Gamma \left[k \left(\alpha_{t} + x_{t}\right)\right]} \IEEEnonumber \\
& \,\,\, \times \exp \! \left[-\theta_{t+1}k \left(\beta_{t} + 1 \right)\right] \! \left(\theta_{t+1}\right)^{k\left(\alpha_{t}+x_{t}\right)-1} \!\!\!\!\!\! . 
\end{IEEEeqnarray}

(\ref{after1}) and (\ref{after2}) mean that the transformed distribution of $\theta_{t+1}$ becomes the Gamma distribution with the following parameters:
\begin{eqnarray}
\label{recursive alpha}
\left\{
\begin{array}{l}
\alpha_{t+1} = k\left(\alpha_{t}+x_{t}\right) \,; \\
\beta_{t+1} =  k\left(\beta_{t}+1\right) \,.
\end{array}
\right.
\end{eqnarray}

If (\ref{recursive alpha}) is recursively applied with respect to $t$, the following equations are obtained:
\begin{eqnarray}
\label{recursive2}
\left\{
\begin{array}{l}
\alpha_{t+1} = k^{t}\alpha_{1}+\sum_{i=1}^{t}k^{t+1-i}x_{i} \,; \\
\beta_{t+1} = k^{t}\beta_{1}+\sum_{i=1}^{t}k^{i-1} \,. 
\end{array}
\right.
\end{eqnarray}

The above equations contribute drastic reduction of calculation complexity.


\subsection{Output Estimator $\hat{x}_{t+1}$}
\label{output}

The output of proposed model is an estimator $\hat{x}_{t+1}$ as depicted in Figure \ref{Model}. $\hat{x}_{t+1}$ is a prediction of number of request arrivals at time $(t+1)$ under the input sequence $x_{1}^{t}=x_{1}x_{2}\cdots x_{t}$. $\hat{x}_{t+1}$ is formulated in terms of statistical decision theory \cite{Berger}\cite{Bernardo} and derived as the following.

Let $\hat{x}_{t+1}$ be an estimator of $x_{t+1}$ and define the following squared-error loss function to evaluate $\hat{x}_{t+1}$:
\begin{IEEEeqnarray}{Cc}
\label{loss}
L(\hat{x}_{t+1}, x_{t+1}) &= (\hat{x}_{t+1}-x_{t+1})^{2} .
\end{IEEEeqnarray}
Since $x_{t+1}$ distributes with the Poisson defined in (\ref{poisson}), {\it risk function} \cite{Berger}\cite{Bernardo} under the certain density parameter $\theta_{t+1}$ becomes the following:
\begin{IEEEeqnarray}{Lls}
\label{risk}
\!\!\!\!\!\!\!\!\!\! R(\hat{x}_{t+1}, \theta_{t+1}) \, &= \!\!\!\!\! \sum_{x_{t+1}=0}^{\infty} \!\!\! \left[ L(\hat{x}_{t+1}, x_{t+1})\, p(x_{t+1} \! \bigm| \theta_{t+1}) \right]\\
\, &= \!\!\!\!\! \sum_{x_{t+1}=0}^{\infty} \!\!\! \left[(\hat{x}_{t+1}-x_{t+1})^{2}\, p(x_{t+1} \!\! \bigm| \! \theta_{t+1})\right] .
\end{IEEEeqnarray}
Next, {\it the Bayes risk function} \cite{Berger}\cite{Bernardo}, which is obtained by taking expectation of the risk function with respect to density parameter $\theta_{t+1}$, becomes the following:
\begin{IEEEeqnarray}{Ll}
\!\!\!\!\!\! \lefteqn{BR(\hat{x}_{t+1})} \IEEEnonumber \\
\!\!\!\!\!\! &= \!\! \int_{0}^{\infty} \!\!\!\! R(\hat{x}_{t+1}, \theta_{t+1}) p(\theta_{t+1}) d\theta_{t+1}\\
\label{bayes risk}
\!\!\!\!\! &= \!\!\!\!\! \sum_{x_{t+1}=0}^{\infty} \!\! (\hat{x}_{t+1}-x_{t+1})^{2} \!\! \int_{0}^{\infty} \!\!\!\!\!\! p(x_{t+1} \!\! \bigm| \! \theta_{t+1}) p(\theta_{t+1}) d\theta_{t+1}\,. 
\end{IEEEeqnarray}
Finally, suppose an estimator $\hat{x}_{t+1}^{*}$ to minimize the Bayes risk function defined in (\ref{bayes risk}). Such estimator is called \textit{the Bayes optimal prediction} \cite{Berger}\cite{Bernardo}. Under this paper's assumptions, $\hat{x}^{*}_{t+1}$ is obtained as follows:
\begin{IEEEeqnarray}{Ll}
\label{expect1}
\!\!\! \lefteqn{\hat{x}_{t+1}^{*}} \IEEEnonumber \\
&= \text{arg} \min_{\hat{x}_{t+1}} BR\left( \hat{x}_{t+1} \right) \\
&= \!\!\! \sum_{x_{t+1}=0}^{\infty} \!\!\!\! x_{t+1} \! \int_{0}^{\infty} \!\!\!\!\!\! p(x_{t+1}\bigm|\theta_{t+1},x_{1}^{t})p(\theta_{t+1} \!\! \bigm| \!\! x_{1}^{t})d\theta_{t+1} \\
\label{expect2}
&= E\left[x_{t+1}\bigm|x_{1}^{t}\right] \\
\label{expect3}
&= E\left[\theta_{t+1} \bigm| x_{1}^{t} \right] \\
\label{expect4}
&= \frac{\alpha_{t+1}}{\beta_{t+1}} \\
\label{prediction2}
&= \frac{k^{t}\alpha_{1}+\sum_{i=1}^{t}k^{t+1-i}x_{i}}{k^{t}\beta_{1}+\sum_{i=1}^{t}k^{i-1}} .
\end{IEEEeqnarray}
Note that (\ref{expect3}) is obtained since $x_{t+1}$ is the Poisson distribution defined in (\ref{poisson}) and the expectation of $x_{t+1}$ corresponds to the that of density parameter $\theta_{t+1}$. (\ref{expect4}) is obtained since $\theta_{t+1}$ has the Gamma distribution defined in (\ref{theta}) and its expectation becomes $\alpha_{t+1}/\beta_{t+1}$. (\ref{prediction2}) is obtained by applying (\ref{recursive2}) to (\ref{expect4}).

\begin{rem}
\label{remark2 of model}

In (\ref{prediction2}), $\hat{x}_{t+1}^{*}$ is obtained by simple arithmetic calculation. This point can be effective not only theoretical point of view but also the real implementation such as server log analysis software tools. The second term of numerator in (\ref{prediction2}) has a form of \textit{Exponentially Weighted Moving Average} \cite{Smith} with a time varying constant $k$. As $k$ becomes larger in (\ref{prediction2}), the weighting of past observed sequence $x_{1}^{t}$ increases. This means that $k$ can be considered as a parameter of long-range dependence (LRD). If $k=1$, its weighting becomes maximum and the model is regarded as stationary model. The effect of $k$ under the real WWW traffic data would be considered in the next section.
\end{rem}



\section{Analysis Examples of WWW Traffic Data}
\label{Simulation}

\subsection{WWW Traffic Data}

The real WWW traffic data was derived from access logs of two different WWW servers (A, and B) on campus. The number of request arrivals was counted with every five-minutes-interval except for maintenance period. The detail specifications are described in Table \ref{data table}.
\begin{table}[h]
\caption{Specifications of real WWW traffic data.}
\label{data table}
\begin{center}
\begin{tabular}{l c c} \toprule
    & Server A & Server B \\ \midrule
 Request Arrivals & 154,932 & 274,302 \\ 
 Start Date & Mar.\ 18, 2005 & Mar.\ 18, 2006 \\ 
 End Date   & Apr.\ 9, 2005 & Apr.\ 9, 2006 \\ 
 Time Length & 13d 12h 10m & 13d 1h 10m \\ 
 Time Intervals & 3,890 & 3,758 \\ \bottomrule
\end{tabular}
\end{center}
\end{table}


\subsection{Maximum Likelihood Estimation for $k$}
\label{MLE Example}

Properties described in the previous section hold on condition that a time varying constant $k$ in (\ref{theta_update}) is known. If real data is dealt with, however, $k$ is unknown and should be estimated. Taking the maximum likelihood estimation of $k$, the objective likelihood function $L(k)$ becomes the following:
\arraycolsep=3pt
\begin{IEEEeqnarray}{Ll}
\lefteqn{L(k)} \IEEEnonumber \\
&= p(x_{1}\bigm|\theta_{1}) \prod_{i=2}^{t} p(x_{i}\bigm|x_{1}^{i-1}, k) \\
&= p(x_{1}\bigm|\theta_{1}) \prod_{i=2}^{t} \! \left[ \int_{0}^{\infty} \!\!\!\!\!\! p(x_{i} \!\! \bigm| \! x_{1}^{i-1} \!\! , \theta_{i}) p(\theta_{i}\! \bigm| \! x_{1}^{i-1})d\theta_{i} \right] \\
\label{likelihood}
&= p(x_{1}\bigm|\theta_{1}) \IEEEnonumber \\
& \times \! \prod_{i=2}^{t} \! \left[ \frac{(\beta_{i})^{\alpha_{i}}\Gamma(\alpha_{i}+x_{i})} {(\beta_{i}+1)^{\alpha_{i}+x_{i}}\Gamma(\alpha_{i})x_{i}!} \! \Biggm | \!\! {\scriptsize \begin{array}{l} \alpha_{i} = k^{i-1}\alpha_{1} \! + \sum^{i-1}_{j=1}k^{i-j}x_{j} \\
\\
\! \beta_{i} = k^{i-1}\beta_{1} + \sum^{i-1}_{j=1}k^{j-1} \\
\end{array} } \!\! \right] . \IEEEnonumber \\
&
\end{IEEEeqnarray}

Note that $\alpha_{i}, \beta_{i}$ in (\ref{likelihood}), the previously obtained results in (\ref{recursive2}) were applied. Therefore, the maximum likelihood estimator (MLE) of $\hat{k}$ is obtained as follows:
\begin{IEEEeqnarray}{Cc}
\label{MLE Equation}
\hat{k} &= \arg \max_{k} L\left(k \right) .
\end{IEEEeqnarray}

\begin{figure}
  \begin{center}
    \begin{tabular}{ll}
      \hspace{-4mm}\resizebox{44mm}{!}{\includegraphics{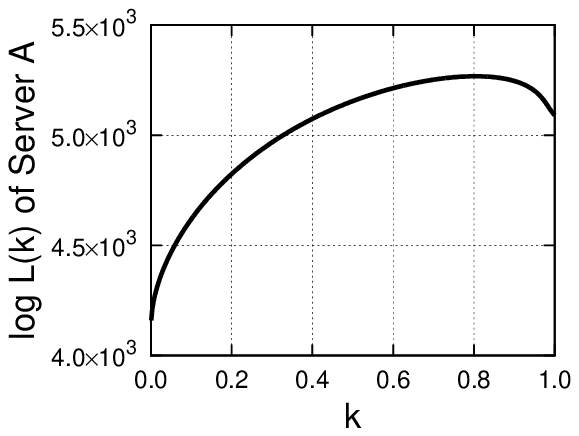}}& \hspace{-8mm}
      \resizebox{45mm}{!}{\includegraphics{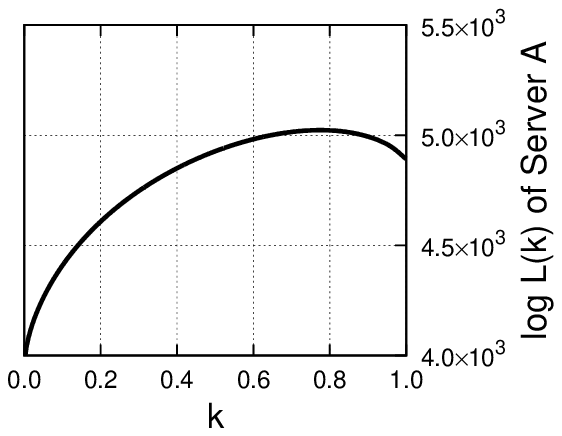}} \\
      \hspace{-4mm}\resizebox{44mm}{!}{\includegraphics{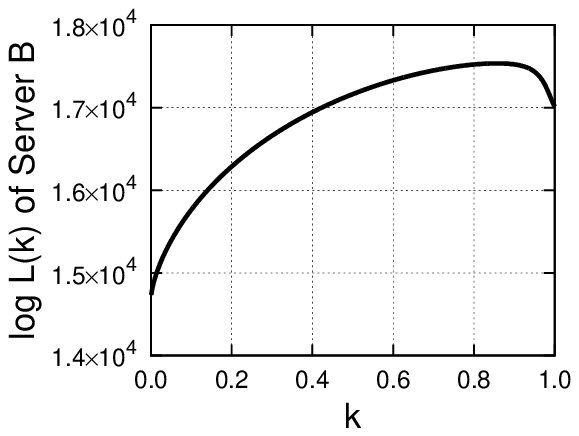}}& \hspace{-8mm}
      \resizebox{45mm}{!}{\includegraphics{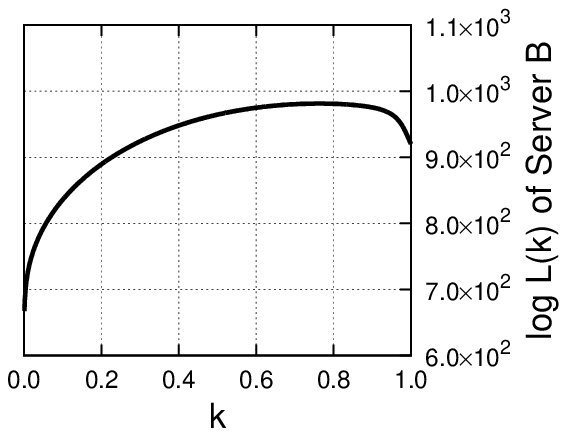}} \\
     \end{tabular}
\caption{Examples of log-likelihood functions. Left top is in Mar.\ 23, 2005 on server A; Right top is in Mar.\ 31, 2005 on server A; Left bottom is in Mar.\ 27, 2006 on server B; Right bottom is in Apr.\ 08, 2006 on server B.}
\label{log-likelihood}
\end{center}
\end{figure}

On the above likelihood function, the analytical maximum likelihood estimation for $k$ is quite difficult, however, the solution can be obtained by numerical calculation. The interval $0 \leq k \leq 1$ is divided into 1,000 sub-intervals and value of $\log L(k)$ is calculated for each $k$ numerically. Some plots of function $\log L(k)$ are shown in Figure \ref{log-likelihood}. Some examples of MLE for $k$ are also shown in Table \ref{MLE Table}.
\begin{table}[h]
\caption{Maximum likelihood estimates for $k$.}
\label{MLE Table}
\begin{center}
\begin{tabular}{c c c} \toprule
  Name  & Date & MLE of $\hat{k}$ \\ \midrule
 Server A & Mar.\ 23, 2005 & 0.804 \\ 
 Server A & Mar.\ 31, 2005 & 0.775 \\ 
 Server B & Mar.\ 27, 2006 & 0.857 \\ 
 Server B & Apr.\ 08, 2006 & 0.764 \\ \bottomrule
\end{tabular}
\end{center}
\end{table}

\subsection{Point Estimation for WWW Traffic Forecasting}

The real WWW data described on Table \ref{data table} was processed to evaluate the point estimates of future request arrivals on the proposed model. For the performance comparison, the point estimates on the classical stationary Poisson model were also calculated. On the proposed model, each MLE of $\hat{k}$ was calculated from the previous day's log. To evaluate performance on both models, the mean squared error between each point estimate and observed value of request arrivals was calculated.

\begin{figure}[h]
\begin{center}
\includegraphics[width=\columnwidth]{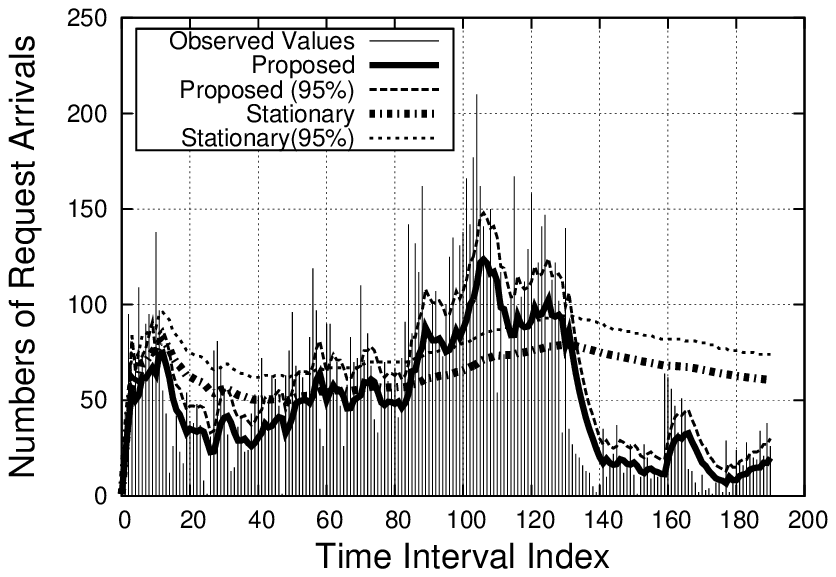}
\caption{Point and interval estimates v.s. observed values plot of server A on Mar.\ 25, 2005 ($\hat{k}=0.804)$.}
\label{Server A1}
\end{center}
\end{figure}

\begin{figure}[h]
\begin{center}
\includegraphics[width=\columnwidth]{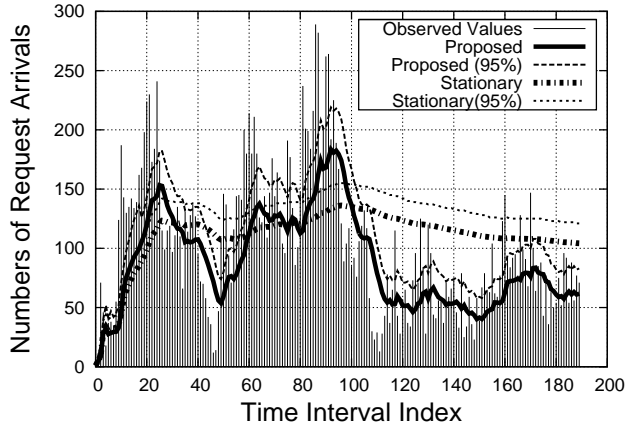}
\caption{Point and interval estimates v.s. observed values plot of server B on Mar.\ 28, 2006 ($\hat{k}=0.857)$.}
\label{Server B1}
\end{center}
\end{figure}

Table \ref{MSE Table A} and \ref{MSE Table B} show mean squared error of proposed and stationary models on server A and B, respectively. In each table, the MLEs of $\hat{k}$ from the previous day's logs are showed on the third row.  Figure \ref{Server A1} shows point and interval estimates v.s. observed values plot of server A on Mar.\ 25, 2005 where $\hat{k}=0.804$. In Figure \ref{Server A1}, the vertical axis is the number of request arrivals and the horizontal axis is time interval index. The solid line, solid chain line, and histogram represent the point estimates on the proposed model, those on the classical stationary Poisson model, and observed values of request arrivals, respectively. Dotted lines are 95\% interval estimates of proposed and stationary models. Figure \ref{Server B1} also shows point and interval estimates v.s. observed values plot of server B on Mar.\ 28, 2006 where $\hat{k}=0.857$.

\begin{table}[h]
\begin{center}
\caption{Mean squared error on server A.}
\label{MSE Table A}
\begin{tabular}{l c c c} \toprule
  Server A   & Proposed Model & $\hat{k}$ & Stationary Model \\ \midrule
 Mar.\ 19 & $2.829 \times 10^{2}$ & 0.716 & $4.988 \times 10^{2}$ \\
 Mar.\ 20 & $2.657 \times 10^{2}$ & 0.762 & $3.297 \times 10^{2}$ \\ 
 Mar.\ 21 & $3.816 \times 10^{2}$ & 0.753 & $4.802 \times 10^{2}$ \\ 
 Mar.\ 22 & $7.111 \times 10^{2}$ & 0.805 & $9.534 \times 10^{2}$ \\ 
 Mar.\ 23 & $8.202 \times 10^{2}$ & 0.759 & $1.335 \times 10^{3}$ \\ 
 Mar.\ 24 & $1.356 \times 10^{3}$ & 0.804 & $3.458 \times 10^{3}$ \\ 
 Mar.\ 25 & $9.523 \times 10^{2}$ & 0.804 & $2.062 \times 10^{3}$ \\ 
 Mar.\ 26 & $4.811 \times 10^{2}$ & 0.783 & $6.479 \times 10^{2}$ \\ 
 Mar.\ 27 & $8.596 \times 10^{2}$ & 0.771 & $1.239 \times 10^{3}$ \\ 
 Mar.\ 28 & $1.980 \times 10^{3}$ & 0.754 & $4.041 \times 10^{3}$ \\ 
 Mar.\ 29 & $1.657 \times 10^{3}$ & 0.777 & $4.019 \times 10^{3}$ \\ 
 Mar.\ 30 & $4.940 \times 10^{2}$ & 0.788 & $7.568 \times 10^{2}$ \\ 
 Mar.\ 31 & $8.088 \times 10^{2}$ & 0.787 & $1.218 \times 10^{3}$ \\ 
 Apr.\ 01 & $8.967 \times 10^{2}$ & 0.775 & $1.887 \times 10^{3}$ \\ 
 Apr.\ 02 & $1.258 \times 10^{1}$ & 0.753 & $1.220 \times 10^{1}$ \\ 
 Apr.\ 03 & $4.184 \times 10^{0}$ & 0.826 & $4.375 \times 10^{0}$ \\ 
 Apr.\ 04 & $4.206 \times 10^{1}$ & 0.914 & $4.317 \times 10^{1}$ \\ 
 Apr.\ 05 & $4.095 \times 10^{1}$ & 0.666 & $3.808 \times 10^{1}$ \\ 
 Apr.\ 06 & $3.612 \times 10^{1}$ & 0.710 & $4.723 \times 10^{1}$ \\ 
 Apr.\ 07 & $2.813 \times 10^{2}$ & 0.661 & $5.183 \times 10^{2}$ \\ 
 Apr.\ 08 & $2.295 \times 10^{1}$ & 0.786 & $2.325 \times 10^{1}$ \\ 
 Apr.\ 09 & $7.589 \times 10^{0}$ & 0.803 & $8.127 \times 10^{0}$ \\ \bottomrule
\end{tabular}
\end{center}
\end{table}
\begin{table}[h]
\begin{center}
\caption{Mean squared error on server B.}
\label{MSE Table B}
\begin{tabular}{l c c c} \toprule
  Server B   & Proposed Model & $\hat{k}$ & Stationary Model \\ \midrule
 Mar.\ 19 & $3.070 \times 10^{2}$ & 0.777 & $3.874 \times 10^{2}$ \\ 
 Mar.\ 20 & $1.302 \times 10^{3}$ & 0.820 & $2.386 \times 10^{3}$ \\ 
 Mar.\ 21 & $5.159 \times 10^{2}$ & 0.782 & $5.896 \times 10^{2}$ \\ 
 Mar.\ 22 & $1.244 \times 10^{3}$ & 0.832 & $1.650 \times 10^{3}$ \\ 
 Mar.\ 23 & $1.741 \times 10^{3}$ & 0.827 & $4.151 \times 10^{3}$ \\ 
 Mar.\ 24 & $1.109 \times 10^{3}$ & 0.836 & $1.980 \times 10^{3}$ \\ 
 Mar.\ 25 & $4.705 \times 10^{2}$ & 0.812 & $5.436 \times 10^{2}$ \\ 
 Mar.\ 26 & $1.504 \times 10^{3}$ & 0.755 & $1.915 \times 10^{3}$ \\ 
 Mar.\ 27 & $2.964 \times 10^{3}$ & 0.800 & $7.918 \times 10^{3}$ \\ 
 Mar.\ 28 & $1.109 \times 10^{1}$ & 0.857 & $1.932 \times 10^{1}$ \\ 
 Mar.\ 29 & $2.019 \times 10^{3}$ & 0.805 & $5.576 \times 10^{3}$ \\ 
 Mar.\ 30 & $4.370 \times 10^{2}$ & 0.784 & $6.708 \times 10^{2}$ \\ 
 Mar.\ 31 & $5.318 \times 10^{2}$ & 0.799 & $6.750 \times 10^{2}$ \\ 
 Apr.\ 01 & $2.568 \times 10^{2}$ & 0.710 & $3.327 \times 10^{2}$ \\ 
 Apr.\ 02 & $4.739 \times 10^{2}$ & 0.640 & $5.118 \times 10^{2}$ \\ 
 Apr.\ 03 & $6.019 \times 10^{2}$ & 0.745 & $8.237 \times 10^{2}$ \\ 
 Apr.\ 04 & $1.544 \times 10^{3}$ & 0.791 & $5.448 \times 10^{3}$ \\ 
 Apr.\ 05 & $1.449 \times 10^{2}$ & 0.820 & $1.709 \times 10^{2}$ \\ 
 Apr.\ 06 & $4.712 \times 10^{2}$ & 0.791 & $5.620 \times 10^{2}$ \\ 
 Apr.\ 07 & $2.784 \times 10^{2}$ & 0.777 & $3.567 \times 10^{2}$ \\ 
 Apr.\ 08 & $3.200 \times 10^{3}$ & 0.764 & $1.082 \times 10^{4}$ \\ 
 Apr.\ 09 & $1.333 \times 10^{3}$ & 0.841 & $3.128 \times 10^{3}$ \\ \bottomrule
\end{tabular}
\end{center}
\end{table}


\subsection{Interval Estimation for WWW Traffic Forecasting}

Table \ref{Interval Estimation A} shows interval estimation example of server A on Mar.\ 25, 2005. In Table \ref{Interval Estimation A}, $t=104$ is taken since it gives max observed value $x_{104}=210$ as the numbers of request arrivals. The second, third, and forth rows show the expected value, 95\% confidence limit, and 99\% confidence limit, respectively on the proposed and stationary models. Table \ref{Interval Estimation B} also shows interval estimation result of server B on Mar.\ 28, 2006 where $t=86$ is taken.  

\begin{table}[h]
\begin{center}
\caption{Interval estimation of server A on Mar.\ 25, 2005.}
\label{Interval Estimation A}
\begin{tabular}{l c c} \toprule
 $t=104$ on Server A     & Proposed & Stationary \\ \midrule
 $\hat{x}_{104}$ (Expected Value) & 111 & 69 \\ 
 $\hat{x}_{104}$ (95\% Confidence Limit) & 133 & 83 \\ 
 $\hat{x}_{104}$ (99\% Confidence Limit) & 142 & 89 \\ 
 $x_{104}$ (Max Observed Value)& \multicolumn{2}{c}{210} \\ \bottomrule
\end{tabular}
\end{center}
\end{table}
\begin{table}[h]
\begin{center}
\caption{Interval estimation of server B on Mar.\ 28, 2006.}
\label{Interval Estimation B}
\begin{tabular}{l c c} \toprule
 $t=86$ on Server B     & Proposed & Stationary \\ \midrule
 $\hat{x}_{86}$ (Expected Value) & 148 & 126 \\
 $\hat{x}_{86}$ (95\% Confidence Limit) & 170 & 145 \\
 $\hat{x}_{86}$ (99\% Confidence Limit) & 180 & 153 \\
 $x_{86}$ (Max Observed Value)& \multicolumn{2}{c}{289} \\ \bottomrule
\end{tabular}
\end{center}
\end{table}

\section{Discussion}
\label{Discussion}

\subsection{Maximum Likelihood Estimation for $k$}
\label{Maximum Likelihood Estimation for k}

According to Figure \ref{log-likelihood}, it is showed that there exist some cases where their likelihood functions of $\log L(k)$ are convex. Actually, all likelihood functions were convex to the best of numerical calculations in this paper. 
Figure \ref{log-likelihood} also shows that the absolute value of gradient in $\log L(k)$ around MLE of $\hat{k}$ becomes quickly larger as $k$ increases beyond $\hat{k}$. This fact shows that the under estimation for $k$ causes less error than its over estimation.

For the estimated value of $k$, the minimum was $\hat{k}=0.640$ and the maximum was $\hat{k}=0.914$ as shown in Table \ref{MLE Table}, \ref{MSE Table A}, and \ref{MSE Table B}. Most of $\hat{k}$ distributes between $0.700$ and $0.850$. This fact suggests non-stationarity of the WWW traffic data. Table \ref{AIC on Server A} shows Akaike Information Criterion (AIC) on server A. According to Table \ref{AIC on Server A}, most of AIC on the proposed model are smaller than those of stationary model to select the proposed model. Exception is that absolute values of AIC on the two models suddenly become smaller around Apr.\ 01, 2005 on server A. This is actually because the average traffic on server A drastically decreased to less than 1,000 request arrivals per a day. In this period, the traffic in each time interval stayed at lower level and could be regarded as the stationary Poisson model. For server B, on the other hand, such traffic decrease did not occur and the proposed model is chosen for all days by AIC model selection. As a whole, it can be concluded that the proposed model has stronger validity for real WWW traffic data than the stationary model from the viewpoint of model selection. 

\subsection{Point Estimation for WWW Traffic Forecasting}

For point estimation of future request arrivals, Table \ref{MSE Table A}, and \ref{MSE Table B} show that the proposed model has the better performance than that of the stationary model in terms of mean squared error. Figure \ref{Server A1} and \ref{Server B1} also depict that the point estimates on the proposed model are following more closely to the observed values than those on stationary model. As mentioned in \textit{Remarks \ref{remark1 of model}}, the proposed model contains the stationary model as a special case when $k=1.000$. Therefore, regardless of its stationarity or non-stationarity of WWW traffic, the proposed model can be applied to traffic forecasting and would help for planning of WWW servers to some extent.

\begin{figure}[h]
\begin{center}
\includegraphics[width=\columnwidth]{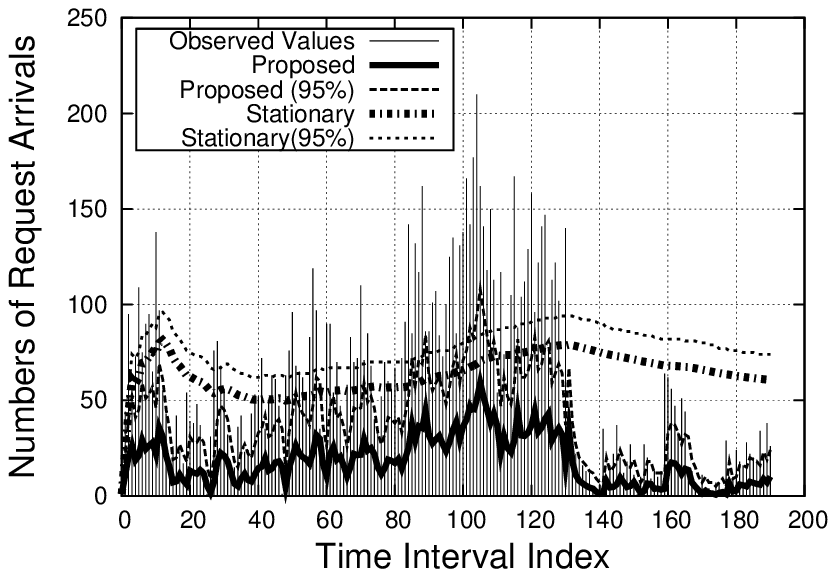}
\caption{Point and interval estimates v.s. observed values plot of server A on Mar.\ 25, 2005 ($k=0.300)$.}
\label{Server A2}
\end{center}
\end{figure}

\begin{figure}[h]
\begin{center}
\includegraphics[width=\columnwidth]{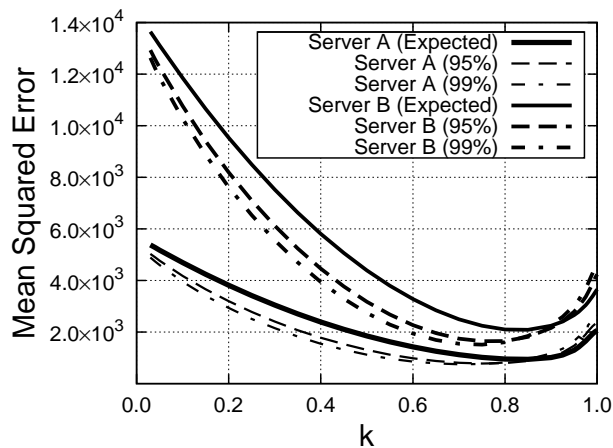}
\caption{Mean squared error v.s. $k$ plot of server A on Mar.\ 25, 2005 and server B on Mar.\ 28, 2006.}
\label{MSE v.s. k}
\end{center}
\end{figure}

However, it should be noted that this result strongly depends on the accuracy of MLE of $k$. In Table \ref{MSE Table A}, each MLE of $k$ during days in April often differs from $\hat{k}=1.000$ in spite of its strong stationarity on the real traffic. In such situation, the mean squared error on the proposed model becomes larger than that of stationary model. Another example is that if $k=0.300$ on the proposed model with server A, the mean squared error of the proposed model becomes $5.41 \times 10^{3}$ where that of stationary model becomes $3.46 \times 10^{3}$. Figure \ref{Server A2} depicts this poor performance of the proposed model. Figure \ref{MSE v.s. k} is a plot of mean squared errors v.s. $k$. In interval of $k < 0.600$, the extremely smaller estimate of $k$ could cause larger mean squared error. The under estimation near the MSE minimizer of $k$, however, causes relatively smaller error than its over estimation as previously described in subsection \ref{Maximum Likelihood Estimation for k}.

In Figure \ref{MSE v.s. k}, mean squared errors takes minimum values around $k=0.800$. In fact, Table \ref{MSE Table A} and \ref{MSE Table B} show that corresponding maximum likelihood estimates for $k$ are $k=0.783$ for server A and $k=0.805$ for server B, respectively. This result suggests that the data length of previous day's log at the maximum likelihood estimation for $k$ was sufficient.

\subsection{Interval Estimation for WWW Traffic Forecasting}

For interval estimation, time interval indices that give the maximum number of request arrivals are taken on Table \ref{Interval Estimation A} and \ref{Interval Estimation B}. This is because one of administrators' concerns can be the maximum number of the request arrivals in terms of stable server operations. As a result, each confidence limit of $x_{t}$ derives larger value than that of point estimate of $x_{t}$ (=expected value) and reduces the mean squared error than that of point estimate. This effect on the proposed model would be stronger than that on the stationary model, since the performance of point estimates of $x_{t}$ on the proposed model is superior to that of stationary model. Thus the advantage of Bayesian approach was observed.

\begin{table}[h]
\caption{Akaike Information Criterion (AIC) on server A.}
\label{AIC on Server A}
\begin{center}
\begin{tabular}{c c c} \toprule
 AIC on Server A  & Proposed & Stationary \\ \midrule
 Mar.\ 19, 2005 & $-3.562 \times 10^{3}$ & $-3.367 \times 10^{3}$ \\ 
 Mar.\ 20, 2005 & $-3.102 \times 10^{3}$ & $-3.011 \times 10^{3}$ \\ 
 Mar.\ 21, 2005 & $-5.085 \times 10^{3}$ & $-4.960 \times 10^{3}$ \\ 
 Mar.\ 22, 2005 & $-8.316 \times 10^{3}$ & $-8.108 \times 10^{3}$ \\ 
 Mar.\ 23, 2005 & $-1.052 \times 10^{4}$ & $-1.018 \times 10^{4}$ \\ 
 Mar.\ 24, 2005 & $-1.797 \times 10^{4}$ & $-1.715 \times 10^{4}$ \\ 
 Mar.\ 25, 2005 & $-1.088 \times 10^{4}$ & $-1.031 \times 10^{4}$ \\ 
 Mar.\ 26, 2005 & $-3.917 \times 10^{3}$ & $-4.649 \times 10^{3}$ \\ 
 Mar.\ 27, 2005 & $-9.023 \times 10^{3}$ & $-8.723 \times 10^{3}$ \\ 
 Mar.\ 28, 2005 & $-1.875 \times 10^{4}$ & $-1.795 \times 10^{4}$ \\ 
 Mar.\ 29, 2005 & $-1.869 \times 10^{4}$ & $-1.774 \times 10^{4}$ \\ 
 Mar.\ 30, 2005 & $-5.825 \times 10^{3}$ & $-5.610 \times 10^{3}$ \\ 
 Mar.\ 31, 2005 & $-1.003 \times 10^{4}$ & $-9.784 \times 10^{3}$ \\ 
 Apr.\ 01, 2005 & $-6.456 \times 10^{3}$ & $-5.783 \times 10^{3}$ \\ 
 Apr.\ 02, 2005 & $-1.538 \times 10^{1}$ & $-2.313 \times 10^{1}$ \\ 
 Apr.\ 03, 2005 & $+1.763 \times 10^{1}$ & $+8.860 \times 10^{0}$ \\ 
 Apr.\ 04, 2005 & $-1.655 \times 10^{2}$ & $-1.615 \times 10^{2}$ \\ 
 Apr.\ 05, 2005 & $-1.821 \times 10^{2}$ & $-1.803 \times 10^{2}$ \\ 
 Apr.\ 06, 2005 & $-1.771 \times 10^{2}$ & $-1.516 \times 10^{2}$ \\ 
 Apr.\ 07, 2005 & $-2.826 \times 10^{3}$ & $-2.675 \times 10^{3}$ \\ 
 Apr.\ 08, 2005 & $-1.170 \times 10^{2}$ & $-1.184 \times 10^{2}$ \\ 
 Apr.\ 09, 2005 & $+6.273 \times 10^{0}$ & $+4.767 \times 10^{-1}$ \\ \bottomrule
\end{tabular}
\end{center}
\end{table}
\begin{table}[h]
\caption{Akaike Information Criterion (AIC) on server B.}
\label{AIC on Server B}
\begin{center}
\begin{tabular}{c c c} \toprule
 AIC on Server B  & Proposed & Stationary \\ \midrule
 Mar.\ 19, 2006 & $-3.270 \times 10^{3}$ & $-3.195 \times 10^{3}$ \\ 
 Mar.\ 20, 2006 & $-1.244 \times 10^{4}$ & $-1.201 \times 10^{4}$ \\ 
 Mar.\ 21, 2006 & $-1.245 \times 10^{4}$ & $-1.201 \times 10^{4}$ \\ 
 Mar.\ 22, 2006 & $-1.652 \times 10^{4}$ & $-1.634 \times 10^{4}$ \\ 
 Mar.\ 23, 2006 & $-2.231 \times 10^{4}$ & $-2.165 \times 10^{4}$ \\ 
 Mar.\ 24, 2006 & $-1.217 \times 10^{4}$ & $-1.180 \times 10^{4}$ \\ 
 Mar.\ 25, 2006 & $-4.482 \times 10^{3}$ & $-4.395 \times 10^{3}$ \\ 
 Mar.\ 26, 2006 & $+1.504 \times 10^{3}$ & $+1.915 \times 10^{3}$ \\ 
 Mar.\ 27, 2006 & $-3.504 \times 10^{4}$ & $-3.403 \times 10^{4}$ \\ 
 Mar.\ 28, 2006 & $-2.130 \times 10^{4}$ & $-2.086 \times 10^{4}$ \\ 
 Mar.\ 29, 2006 & $-1.958 \times 10^{4}$ & $-1.852 \times 10^{4}$ \\ 
 Mar.\ 30, 2006 & $-4.601 \times 10^{3}$ & $-4.428 \times 10^{3}$ \\ 
 Mar.\ 31, 2006 & $-3.417 \times 10^{3}$ & $-3.282 \times 10^{3}$ \\ 
 Apr.\ 01, 2006 & $-8.741 \times 10^{2}$ & $-8.145 \times 10^{2}$ \\ 
 Apr.\ 02, 2006 & $-4.393 \times 10^{3}$ & $-4.308 \times 10^{3}$ \\ 
 Apr.\ 03, 2006 & $-7.633 \times 10^{3}$ & $-7.462 \times 10^{3}$ \\ 
 Apr.\ 04, 2006 & $-1.837 \times 10^{4}$ & $-1.710 \times 10^{4}$ \\ 
 Apr.\ 05, 2006 & $-1.033 \times 10^{3}$ & $-9.828 \times 10^{2}$ \\ 
 Apr.\ 06, 2006 & $-3.311 \times 10^{3}$ & $-3.183 \times 10^{3}$ \\ 
 Apr.\ 07, 2006 & $-1.958 \times 10^{3}$ & $-1.838 \times 10^{3}$ \\ 
 Apr.\ 08, 2006 & $-3.384 \times 10^{4}$ & $-3.211 \times 10^{4}$ \\ 
 Apr.\ 09, 2006 & $-1.090 \times 10^{4}$ & $-1.010 \times 10^{4}$ \\ \bottomrule
\end{tabular}
\end{center}
\end{table}

\section{Conclusion}
\label{Conclusion}

This paper showed Bayesian forecasting of WWW traffic on the time varying Poisson model. This model is obtained by defining a random-walk type of time varying parameter function on Simple Power Steady Model. The forecasting estimator of this model guarantees the Bayes optimality in terms of statistical decision theory and is calculated by simple arithmetic calculation. The latter point especially can be effective for the real implementation such as server log analysis software tools. 

Furthermore, the non-stationarity is expressed by a time varying degree constant $k$ in the model. This paper pointed out that the constant $k$ can be considered as a parameter of long-range dependent (LRD) for real traffic data and the model includes stationary Poisson model as a special case if $k=1$. 

For evaluation of Bayesian approach, the real WWW traffic data is applied to the model in this paper. The maximum likelihood estimation method of $k$ from real traffic data is also discussed and its performance is proved to be sufficient for WWW traffic forecasting.

According to its result, the proposed model has stronger validity than classical stationary Poisson model in terms of model selection. Furthermore, under the estimated value of $k$, the point and interval estimates on the proposed model showed smaller mean squared error comparing to those on the stationary model for the traffic forecasting. Thus the advantage of the proposed model is shown from both theoretical and empirical points of view.






\section*{Acknowledgment}

The first and second authors are supported by the following research grants:

\begin{itemize}
\item Grant-in-Aid for Scientific Research on Innovative Areas, Japan Society for the Promotion of Science (No.20200044)

\item Waseda University Grant for Special Research Projects (No.2009A-058)
\end{itemize}


\end{document}